\documentclass[runningheads]{llncs}

\usepackage{graphicx}
\usepackage{amsmath,amssymb,amsfonts}

\usepackage{multirow}
\usepackage{booktabs}
\usepackage{subfigure}
\usepackage{wrapfig}
\usepackage{url}
\usepackage[T1]{fontenc}
\usepackage[misc,geometry]{ifsym}
\usepackage{graphicx}
\begin{document}
\title{AFSC: Adaptive Fourier Space Compression for Anomaly Detection}
%
%
\author{Haote Xu\inst{1} \and
	Yunlong Zhang\inst{2} \and
	Liyan Sun\inst{1} \and
	Chenxin Li\inst{1} \and
	Yue Huang\inst{1} \and
	Xinghao Ding\inst{1}(\Letter) }

\authorrunning{Xu et al.}
%

%
\institute{School of Informatics, Xiamen University, Xiamen, China
	\\
	\email{dxh@xmu.edu.cn} \and
	School of Engineering, Westlake University, Hangzhou, China
}
\maketitle              
\begin{abstract}
Anomaly Detection (AD) on medical images enables a model to recognize any type of anomaly pattern without lesion-specific supervised learning. Data augmentation based methods construct pseudo-healthy images by “pasting” fake lesions on real healthy ones, and a network is trained to predict healthy images in a supervised manner. The lesion can be found by difference between the unhealthy input and pseudo-healthy output. However, using only manually designed fake lesions fail to approximate to irregular real lesions, hence limiting the model generalization. We assume by exploring the intrinsic data property within images, we can distinguish previously unseen lesions from healthy regions in an unhealthy image. In this study, we propose an Adaptive Fourier Space Compression (AFSC) module to distill healthy feature for AD. The compression of both magnitude and phase in frequency domain addresses the hyper intensity and diverse position of lesions. Experimental results on the BraTS and MS-SEG datasets demonstrate an AFSC baseline is able to produce promising detection results, and an AFSC module can be effectively embedded into existing AD methods.

\end{abstract}
\section{Introduction}
Lesion Detection is one of the important tasks in medical image analysis. Deep Learning (DL) has been developed into a powerful tool with a wide gamut of applications, spanning a wide range of medical image analysis applications including detection of lesions. DL based supervised lesion detection methods are virtually ineffective in the face of emergent diseases (e.g. covid-19) that have not been seen in the training phase. In addition, these methods need to meet pathology-sufficiency conditions (with a large number of pathological images in the training phase) and require image-level or pixel-level labels, which are time-consuming and laborious to obtain. Consequently, Anomaly Detection (AD) has logically become a very promising clinical diagnostic aid as a way of dealing with pathology-deficiency situations (lack of pathological images during the training phase)\cite{chen2021normative}.

Since the rise of Auto-Encoders (AEs)\cite{vincent2008extracting}and its applications in AD\cite{baur2021autoencoders}, a plethora of new, AD based on pseudo health generation approaches has appeared. This type of method exploits the network to generate a pseudo-healthy image paired with an anomalous image and locates the anomalous region by the difference between the pair\cite{dey2021asc}. AnoGAN by Schlegl et al \cite{schlegl2017unsupervised}, a network for finding pseudo-healthy images paired with input images from latent space was proposed, by learning a mapping from image space to latent space. However, the inference process of AnoGAN is very time consuming as it needs to iteratively search for the closest pseudo-healthy image in latent space for the input image. Recently, in order to avoid the tedious search process like AnoGAN in the inference phase, several approaches have been dedicated to finding artificially generated anomaly image methods with generalizability. The aim is for models to have the ability to recover artificial anomalous images to normal images and generalize to real anomalous images during the testing phase. For example, Vitjan et al. \cite{zavrtanik2021draem} have proposed a method to generate anomalous images by overlaying randomly sampled texture images, which were obtained from a combination of multiple enhancement functions, onto normal images. 

These methods add out-of-distribution anomalies to the original health image, but the artificial anomaly image will still have distributional differences from the real anomaly image, which will lead to limited generalization ability of the model. Therefore, to capture the most essential information used to generate pseudo-healthy images and make the model more generalizable, we are working to find a subtraction method. In this way, RIAD\cite{zavrtanik2021reconstruction} attempts to randomly mask the image to lose lesion information and subsequently repair the lost anomalous regions to healthy areas by inpainting. However, there are limitations to RIAD due to the stochastic character of the masked area and the nonxed nature of the lesion regions. Therefore, we step outside the image space and turn our view into Fourier space. Each value of the phase and amplitude spectra in Fourier space contains information about the spatial and intensity of the whole image, respectively \cite{jain1989fundamentals}, so we assume that compressing both to capture the most essential health information for generating pseudo-health images will greatly reduce the interference of the lesions during testing. A dilemma is how to effectively compress the Fourier space for AD tasks? To this end, we propose an Adaptive Fourier Space Compression (AFSC) for Anomaly Detection inspired by Monte Carlo strategies in \cite{kingma2013auto}.

\begin{figure}
	\includegraphics[width=\textwidth]{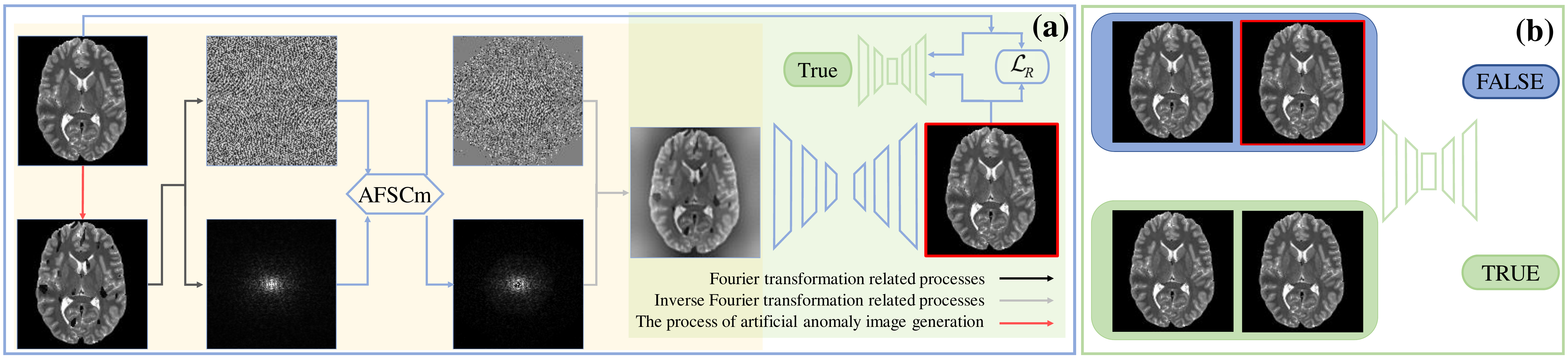}
	\caption{The framework of the proposed AFSC. (a)The main part contains the adaptive Fourier space compression module on a yellow background and the generator module on a green background. (b) The secondary component includes a discriminator which can identify pairs of real images and pseudo-healthy generated images.} \label{fig1}
\end{figure}

In summary, our contributions are: (a) Rather than following the conventional way in which existing methods operate on the image space, we use information compression of phase and amplitude in Fourier space to improve the performance of AD methods based on pseudo-health generation; (b) We propose a flexible adaptive Fourier space compression module for finding the most efficient health region generation information, which is applicable to most pseudo-health generation based AD methods; (c) We have experimentally demonstrated that information compression on Fourier space can bring the performance of some existing pseudo-health based anomaly detection SOTA methods to a higher level, which is promising for AD.

\section{Method}

We show the architecture of the AFSC in Figure \ref{fig1} with an Adaptive Fourier Space Compression module (AFSCm) and Pseudo-health Generation module(PGm). The most critical information in Fourier space for pseudo-health generation is extracted by AFSCm and as input for PGm to generate pseudo-health images. In the following subsections we describe in detail the AFSC.

\subsection{General framework}
An artificial anomaly image $x\in {{\mathbb{R}}^{H\times W}}$ is obtained by adding anomalies to a single-channel health image ${{x}_{\text{ori}}}\in {{\mathbb{R}}^{H\times W}}$, as exemplified with MR T2- weighted image, through existing artificial anomaly generation methods (e.g. \cite{zavrtanik2021draem}). The purpose of this processing is for the image information to be more discriminately extracted by the model and used to generate pseudo-healthy images. The Fourier transformation of $x$ is expressed as:

\begin{equation}
\mathcal{F}(x)(u,v)=\sum\limits_{h=0}^{H-1}{\sum\limits_{w=0}^{W-1}{x}}(h,w){{e}^{-j2\pi }}\left( \frac{h}{H}u+\frac{w}{W}v \right).
\end{equation}

The Fourier transformation of $x$ can further be expressed as its amplitude and phase components by the following equation:

\begin{equation}
\mathcal{A}(x)(u,v)={{\left[{{R}^{2}}(x)(u,v)+{{I}^{2}}(x)(u,v)\right]}^{1/2}},
\end{equation}

\begin{equation}
\mathcal{P}(x)(u,v)=\arctan \left[\frac{I(x)(u,v)}{R(x)(u,v)}\right],
\end{equation}
where $R\left( x \right)$, $I\left( x \right)$ denote the real and imaginary part of $\mathcal{F}(x)$, respectively.

Inspired by the physical properties of Fourier phase and amplitude\cite{jain1989fundamentals}, we attempt to construct models dedicated to extracting the most effective health information in Fourier space for pseudo-health generation. A natural choice for this purpose is to achieve information compression by a form of adaptive masking of the amplitude and phase values in the Fourier space of the original image. Inspired by \cite{bahadir2020deep}, we devise an optimizable phase and amplitude compression strategy. After compression in this way the amplitude and phase become:

\begin{equation}
\widehat{\mathcal{A}}\left( x \right)={{M}_{\mathcal{A}}}\odot \mathcal{A}\left( x \right),
\end{equation}

\begin{equation}
\widehat{\mathcal{P}}\left( x \right)={{M}_{\mathcal{P}}}\odot \mathcal{P}\left( x \right),
\end{equation}
where ${{M}_{\mathcal{A}}},{{M}_{\mathcal{P}}}\in {{\mathbb{R}}^{H\times W}}$ are the masks corresponding to the phase and amplitude obtained by adaptive learning, respectively. The details of ${{M}_{\mathcal{A}}}$ and ${{M}_{\mathcal{P}}}$ will be presented in the next subsection. The amplitude and phase of the information compression are combined to form a new Fourier representation:

\begin{equation}
\mathcal{F}\left( {\overset{\scriptscriptstyle\smile}{x}} \right)(u,v)=\widehat{\mathcal{A}}\left( x \right)(u,v)*{{e}^{-j*\widehat{\mathcal{P}}\left( x \right)(u,v)}}.
\end{equation}

The Fourier-space compressed image $\hat{x}$ is obtained by the Fourier inverse transformation ${{\mathcal{F}}^{-1}}\left[ \mathcal{F}\left( {\overset{\scriptscriptstyle\smile}{x}} \right)(u,v) \right]$, where ${{\mathcal{F}}^{-1}}(\cdot )$ defines the inverse Fourier transformation. Ultimately, $\hat{x}$ is used as input to the generative network for the pseudo-healthy images generated. The generation module in AFSC consists of a generator and a discriminator. The loss function of the generator is as follows:

\begin{equation}\label{eq7}
{{\mathcal{L}}_{G}}=\lambda {{\mathcal{L}}_{R}}+{{\mathcal{L}}_{d1}}+\delta {{\mathcal{L}}_{M}}.
\end{equation}
The first term of Eq.\ref{eq7} $\lambda {{\mathcal{L}}_{R}}=\lambda {{\mathcal{L}}_{mae}}\left( {{x}_{\text{ori}}},\mathbf{G}\left( {\hat{x}} \right) \right)$ is the reconstruction loss, where $\mathbf{G}\left( \cdot  \right)$ is the generator, ${{\mathcal{L}}_{mae}}\left( \cdot  \right)$denotes pixel-wise ${{\mathcal{L}}_{1}}$ loss, and $\lambda $ is a hyperparameter. The second term ${{\mathcal{L}}_{d1}}={{\mathcal{L}}_{mse}}\left( (\mathbf{D}\left( \mathbf{G}\left( {\hat{x}} \right),{{x}_{\text{ori}}} \right),{{y}_{1}} \right)$ is the adversarial loss, where ${{y}_{1}}\in {{\mathbb{R}}^{H\times W}}$ is a matrix with all elements 1, ${{\mathcal{L}}_{mse}}\left( \cdot  \right)$denotes pixel-wise ${{\mathcal{L}}_{2}}$ loss, and $\mathbf{D}\left( \cdot  \right)$ is a discriminator used to discriminate whether two input images are identical at the pixel level. The last term $\delta {{\mathcal{L}}_{M}}=\delta \left( {{\left\| {{M}_{\mathcal{P}}} \right\|}_{1}}+{{\left\| {{M}_{\mathcal{A}}} \right\|}_{1}} \right)$ allows ${{M}_{\mathcal{A}}}$ and ${{M}_{\mathcal{P}}}$ to satisfy the sparsity property. The larger the hyperparameter $\delta $ the more sparse masks can be. To fool the discriminator, the generator will further reduce the distribution difference between the generated image and the real health image. As an adversary to the generator, the loss function of the discriminator is
\begin{equation}\label{eq8}
{{\mathcal{L}}_{d2}}={{\mathcal{L}}_{d2\_2}}+{{\mathcal{L}}_{d2\_1}}.
\end{equation}
Both the first terms ${{\mathcal{L}}_{d2\_1}}={{\mathcal{L}}_{mse}}\left( \mathbf{D}\left( \mathbf{G}\left( {\hat{x}} \right).\text{detach}(),{{x}_{\text{ori}}} \right),{{y}_{0}} \right)$ and second terms ${{\mathcal{L}}_{d2\_2}}={{\mathcal{L}}_{mse}}\left( \mathbf{D}\left( {{x}_{\text{ori}}},{{x}_{\text{ori}}} \right),{{y}_{1}} \right)$ in Eq.\ref{eq8} are designed to enable the discriminator to have the ability to discriminate the difference in pixel-wise between the generated image and original image, where ${{y}_{0}}\in {{\mathbb{R}}^{H\times W}}$ represents a zero matrix and detach() is a gradient-stop operation.

\subsection{Adaptive Fourier space compression module}

For the ease of illustration in the following, two different masks ${{M}_{\mathcal{A}}}$, ${{M}_{\mathcal{P}}}$ are substituted by $M\in {{\mathbb{R}}^{H\times W}}$. We assume that each element ${{m}_{i}}$ of $M$ is obtained by sampling in the Bernoulli distribution with probability ${{p}_{i}}$ denoted ${{m}_{i}}\sim\operatorname{Bern}\left( {{p}_{i}} \right)$. Sampling of the Bernoulli distribution is achieved by reparameterising ${{m}_{i}}=\Delta \left\{ {{u}_{i}}\le {{p}_{i}} \right\}$ in AFSC, where $\Delta \left\{ \cdot  \right\}$ is a function that maps a Boolean-true input to 1 and 0 otherwise, and ${{u}_{i}}$ is a random variable sampled from a [0,1] uniform distribution.

In order to make the whole process optimizable, we first obtain a uniformly distributed $\omega \in {{\mathbb{R}}^{H\times W}}$ with all initial values of elements from [0,1], and map $\omega $ by an inverse function of a sigmoid function ${{\sigma }_{1}}\left( \cdot  \right)$ (with slope 10) to obtain a variable $\overset{\scriptscriptstyle\frown}{\omega }$ that is optimizable in the network . Thus, the probability map $p={{\sigma }_{1}}\left( {\overset{\scriptscriptstyle\frown}{\omega }} \right)$ can be built by $\overset{\scriptscriptstyle\frown}{\omega }$. 

The optimization process is hindered due to the non-differentiability of $\Delta \left\{ \cdot  \right\}$. Hence we relax the function $\Delta \left\{ \cdot  \right\}$ by approximating it as a differentiable sigmoid function ${{\sigma }_{2}}\left( \cdot  \right)$ with slope 12:
\begin{equation}
	{{M}_{s}}={{\sigma }_{2}}\left( {{\sigma }_{1}}(\overset{\scriptscriptstyle\frown}{\omega })-u \right),\quad u\sim U(0,1).
\end{equation}

Ultimately, $M$ with the ability to judge is obtained as the network is continuously optimised which can select the elements of Fourier space best adapted to pseudo-health generation.
\subsection{Implementation}

We obtain two binarization-processed masks with sparsity $s$ by separately sorting the values in ${{M}_{a}}$ and ${{M}_{p}}$ in descending order at the end of the penultimate training period, and setting each to a value of 1 for the first $\left\lfloor s\times HW \right\rfloor $ elements and 0 for the remaining elements. The masks were used to fine-tune the network in the last training period. We implemented the AFSC on a graphic card TITAN Xp with Pytorch as the platform. In particular, AFSCm in our framework can be joined to the majority of existing AD methods based on pseudo-health generation to improve generalization.

\section{Experiments}

\subsection{Datasets and metrics}

We evaluated AFSC using T2-weighted images from three different brain MRI datasets in 2 medical anomaly detection tasks: multiple sclerosis lesion detection, and brain tumour detection. HCP\cite{van2012human} was as training set for all models. We worked with a version of the HCP 500 Subjects Release (S500), which contained 526 healthy subjects. BraTS2017 \cite{menze2014multimodal,bakas2017advancing,bakas2018identifying} was used as one of the test sets with 210 subjects show high-grade glioblastomas and 75 subjects show low-grade gliomas. Another test set, MS-SEG2015\cite{carass2017longitudinal}, has 21 MRI brain image data from 5 subjects with sub-acute ischemic stroke at different times. We normalized the intensity of each subject's T2w image over the 3D volume separately. We trained and tested all axial slices with a resolution resized to $128\times 128$.

In this paper, we evaluate AFSC in terms of image-level anomaly detection and pixel-level anomaly localization. Following \cite{zavrtanik2021draem}, the standard metric of anomaly detection, AUROC, is used in the evaluation of our method. Image-level AUROC is applied to overall slice anomaly detection, and pixel-level AUROC is used for per-slice anomaly localization evaluation. We further report the average precision(AP) measure on the image-level and pixel-level for the false positive rate of the test results.

\subsection{Results}

\begin{table}[thb]
	\centering
	\caption{Evaluation results on the BraTS2017 dataset. The best mean of each metric is shown in \underline{underline}.}\label{tab1}
	\setlength{\tabcolsep}{0.5mm}{
		\begin{tabular}{cccccc}
			\hline
			\multirow{2}{*}{Methos} & \multicolumn{5}{c}{Metrics}                \\
			& AUROC$_\text{img}$ & AP$_\text{img}$ & DICE & AUROC$_\text{pix}$ & AP$_\text{pix}$ \\ \hline
			AnoGAN                  &0.827$\pm $0.028          &0.775$\pm $0.044       &0.314$\pm $0.065      &0.810$\pm $0.052          &0.344$\pm $0.082       \\
			AE                      &0.823$\pm $0.008          &0.791$\pm $0.008       &0.353$\pm $0.005      &0.795$\pm $0.007          &0.383$\pm $0.017       \\
			AE$_\text{w/AFSCm}$               &0.847$\pm $0.004          &0.830$\pm $0.006       &0.431$\pm $0.001      &0.862$\pm $0.001          &0.496$\pm $0.003       \\
			DRAEM                   &0.853$\pm $0.019          &0.792$\pm $0.042       &0.302$\pm $0.023      &0.819$\pm $0.008          &0.263$\pm $0.027       \\
			DRAEM$_\text{w/AFSCm}$            &0.841$\pm $0.002          &0.729$\pm $0.017       & 0.297$\pm $0.028     &0.808$\pm $0.016          &0.210$\pm $0.019       \\
			DRAEM\_G                &0.852$\pm $0.003          &0.835$\pm $0.003       &0.479$\pm $0.005      &0.888$\pm $0.001          &0.509$\pm $0.006       \\
			DRAEM\_G$_\text{w/AFSCm}$         &0.868$\pm $0.007          &0.855$\pm $0.008       &\underline{0.500$\pm $0.009}      &0.893$\pm $0.004          &\underline{0.559$\pm $0.011}       \\
			AFSC$_\text{w/o AFSCm}$           &0.816$\pm $0.003          &0.789$\pm $0.011       &0.378$\pm $0.005      &0.822$\pm $0.004          &0.433$\pm $0.001       \\
			AFSC$_\text{w/ rand mask}$                    &0.860$\pm $0.004          &0.836$\pm $0.003       &0.436$\pm $0.006      &0.879$\pm $0.005          &0.488$\pm $0.015       \\ 
			AFSC                    &\underline{0.877$\pm $0.005}          &\underline{0.858$\pm $0.006}       &0.490$\pm $0.010      &\underline{0.899$\pm $0.005}          &0.520$\pm $0.018       \\			\hline
	\end{tabular}}\label{tab1}
\end{table}

\begin{figure}
	\includegraphics[width=\textwidth]{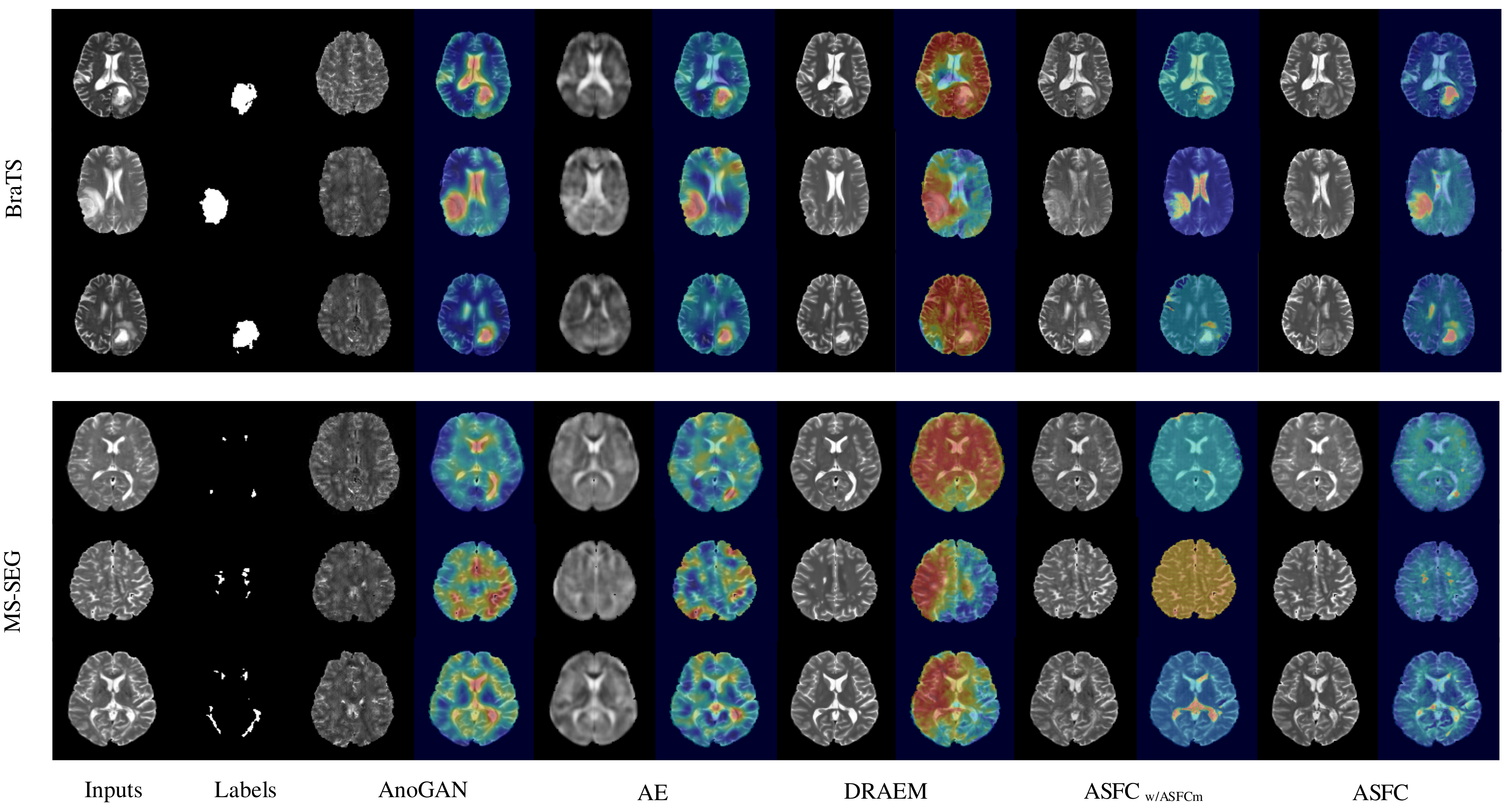}
	\caption{The qualitative results of generated images and anomalous images.The first and second columns show the pathological images and paired lesion annotations.The remaining columns show the corresponding generated images and anomaly localization images for AnoGAN, AE, DRAEM, AFSC$_\text{w/o AFSCm}$ and AFSC respectively.} \label{fig2}
\end{figure}

\begin{table}[thb]
	\centering
	\caption{Experimental results of the proposed method in comparison to baselines on MS-SEG2015.The best mean of each metric is shown in \underline{underline}.}\label{tab2}
	\setlength{\tabcolsep}{0.5mm}{
		\begin{tabular}{cccccc}
			\hline
			\multirow{2}{*}{Methos} & \multicolumn{5}{c}{Metrics}                \\
			& AUROC$_\text{img}$ & AP$_\text{img}$ & DICE & AUROC$_\text{pix}$ & AP$_\text{pix}$ \\ \hline
			AnoGAN                  &0.868$\pm $0.021          &0.785$\pm $0.047       &0.046$\pm $0.009      &0.722$\pm $0.103          &0.022$\pm $0.008       \\
			AE                      &0.867$\pm $0.003          &0.771$\pm $0.010       &0.071$\pm $0.005      &0.787$\pm $0.003          &0.034$\pm $0.002       \\
			AE$_\text{w/AFSCm}$               &0.887$\pm $0.007          &0.801$\pm $0.011       &0.125$\pm $0.003      &0.859$\pm $0.011          &0.066$\pm $0.006       \\
			DRAEM                   &0.894$\pm $0.016          &0.786$\pm $0.032       &0.053$\pm $0.026      &0.685$\pm $0.112          &0.028$\pm $0.018       \\
			DRAEM$_\text{w/AFSCm}$            &\underline{0.912$\pm $0.003}          &\underline{0.822$\pm $0.003}       &0.057$\pm $0.007      &0.784$\pm $0.016          &0.027$\pm $0.008       \\
			DRAEM\_G                &0.837$\pm $0.002          &0.719$\pm $0.011       &0.143$\pm $0.002      &0.892$\pm $0.002          &\underline{0.192$\pm $0.011}       \\
			DRAEM\_G$_\text{w/AFSCm}$	&0.889$\pm $0.018         &0.805$\pm $0.038          &0.176$\pm $0.017       &\underline{0.926$\pm $0.008}      &0.187$\pm $0.009          \\
			AFSC$_\text{w/o AFSCm}$           &0.769$\pm $0.016          &0.631$\pm $0.040      &0.053$\pm $0.001      &0.733$\pm $0.008          &0.031$\pm $0.002       \\
			AFSC$_\text{w/ rand mask}$                    &0.819$\pm $0.008          &0.696$\pm $0.011       &0.161$\pm $0.014      &0.912$\pm $0.007          &0.109$\pm $0.007       \\
			AFSC           &0.814$\pm $0.006          &0.691$\pm $0.008       &\underline{0.184$\pm $0.002}      &0.913$\pm $0.001          &0.143$\pm $0.003       \\ \hline
	\end{tabular}}\label{tab2}
\end{table}

We compared AFSC with other state-of-the-art anomaly detection methods AnoGAN \cite{schlegl2017unsupervised}, DRAEM \cite{zavrtanik2021draem} and AE\cite{baur2021autoencoders}. As DRAEM generates anomaly scores differently from other methods, we additionally obtain anomaly score maps for the pseudo-health images it generates by the same difference calculation as other methods, denoted as DRAEM\_G, for a fair comparison. The joint of AFSCm with AE and DRAEM respectively was used for the comparison of the vanilla methods in order to test the flexibility of AFSCm (AnoGAN cannot be joined with AFSCm, as its input is a random variable rather than an image).  We empirically set the sparsity of the corresponding compression modules to 0.9, 0.5 and 0.5 for AE$_\text{w/AFSCm}$, DRAEM$_\text{w/AFSCm}$ and AFSC respectively (in the experiments of this paper the sparsity settings are the same for phase and amplitude, although different sparsities can be set).

Table \ref{tab1} summarises the results of all methods at BraTS2017. Overall, AFSC achieved optimal results for almost all metrics, compared with other methods. Importantly, all metrics of AE$_\text{w/AFSCm}$,DRAEM\_G$_\text{w/AFSCm}$ have been improved after the joint AFSCm. In particular, DRAEM\_G$_\text{w/AFSCm}$ achieved the best for all methods on the Brats dataset in both dice and APpix metrics. As shown in the visualisation of the results in Figure \ref{fig2}, AFSCm enabled all generated pseudo-healthy images to be more anatomically accurate and made most lesions resemble healthy tissue. Furthermore, we observe that AE$_\text{w/AFSCm}$ performs better than AE, although the input of AE are healthy images. That AFSCm is valid for anomaly detection methods based on pseudo-health generation in which the input are healthy images can still be verified by this phenomenon. Due to the cross-dataset testing, AnoGAN did not use the test samples as input to the generator and only searched for representations of healthy samples in the latent space which were similar to the test images, hence the poor overall test performance.

The experimental results for all methods in Table \ref{tab2} show that AFSC is optimal with the DICE metric on MS-SEG2015. More importantly, the adaptive Fourier space compression by our proposed method makes DRAEM$_\text{w/AFSCm}$ and DRAEM\_G$_\text{w/AFSCm}$ optimal for the almost all remaining metrics. However, the results of all methods were unsatisfactory, due to the fact that MS-SEG2015 lesions are challenging to detect on T2-weighted images. The regions of the lesion are small and not obvious for the T2-weighted images as can be seen in the visualization of the results in Figure \ref{fig2}.

To verify the efficiency of the adaptively obtained masks, we replaced AFSCm in AFSC with a fixed random mask of the same sparsity is tested on BraTS2017 and MS-SEG2015. The comparison results are summarised in Table \ref{tab1} and Table \ref{tab2}, where the original AFSC outperforms the randomly generated masks for the almost all evaluation metrics, which validates that the masks obtained through AFSCm optimization are discriminative for critical health information in terms of pseudo-health generation.

\section{Conclusions}

In this paper, we propose an anomaly detection framework, named AFSC, targeting the extraction of critical information for pseudo-health generation through adaptive compression of Fourier space information to improve model generalization. In particular, AFSCm in AFSC joint with the majority of existing anomaly detection methods based on pseudo-health generation enables these methods to achieve more competitive results was experimentally demonstrated. However, the current version of AFSC is still empirically tuned for sparsity (with smaller sparsity for powerful generative models), which will be the topic of subsequent optimization. Overall, we have compressed information in Fourier space based on the physical properties of phase and amplitude components, which provides a different way of thinking for anomaly detection, and how to make better use of the information in Fourier space in the future is a direction we will explore subsequently.

%
%
%
%

\bibliographystyle{splncs04}
\bibliography{AFSC}

\end{document}